\documentstyle[multicol,psfig,prb,aps]{revtex}
\def\v#1{\mbox{\boldmath$#1$\unboldmath}}
\begin{document}
\draft

\title{Elementary excitations in 
one-dimensional spin-orbital models: \protect\\
neutral and charged solitons and their bound states}

\author{A. K. Kolezhuk\cite{leave} and H.-J. Mikeska}
\address{Institut f\"ur Theoretische Physik, Universit\"at Hannover,
Appelstr. 2, D-30167 Hannover, Germany}
\author{U. Schollw\"ock}
\address{Sektion Physik, Ludwig-Maximilians Universit\"at M\"unchen,
 Theresienstr. 37, D-80333 M\"unchen, Germany}

\date{August 22, 2000}

\maketitle

\begin{abstract}

We study, both numerically and variationally, the interplay between
different types of elementary excitations in the model of a spin chain
with anisotropic spin-orbit coupling, in the vicinity of the ``dimer
line'' with an exactly known dimerized ground state. Our variational
treatment is found to be in a qualitative agreement with the exact
diagonalization results.  Soliton pairs are shown to be the lowest
excitations only in a very narrow region of the phase diagram near the
dimer line, and the phase transitions are always governed by
magnon-type excitations which can be viewed as soliton-antisoliton
bound states.  It is shown that when the anisotropy exceeds certain
critical value, a new phase boundary appears.  In the doped model on
the dimer line, the exact elementary charge excitation is shown to be
a hole bound to a soliton. Bound states of those ``charged solitons'' are
studied; exact solutions for $N$-hole bound states are presented.

\end{abstract}


\begin{multicols}{2}

\section{Introduction}
\label{sec:intro}

One-dimensional systems with coupled spin and orbital degrees of
freedom have been attracting a considerable attention during the last
two years. This interest is partly motivated by the progress in the
experimental study of the quasi-one-dimensional spin-gap materials
$\rm Na_{2}Ti_{2}Sb_{2}O$ and $\rm NaV_{2}O_{5}$, which are believed
to be described by a two-band orbitally degenerate Hubbard model at
quarter filling.\cite{NaTiSbO,NaVO} Orbital degrees of freedom may be
described with the help of pseudospin-1/2 variables; the corresponding
effective Hamiltonian for the two-band Hubbard model was derived long
ago by Kugel and Khomskii \cite{Kugel-Khomskii} and has the following
form:
\begin{eqnarray} 
\label{ham} 
\widehat{H}&=&K\sum_{n}
(\v{S}_{n}\v{S}_{n+1})(\v{\tau}_{n}\v{\tau}_{n+1}
+\varepsilon' \tau_{n}^{z}\tau_{n+1}^{z})\nonumber\\
&+&\sum_{n}\big\{ J_{s}(\v{S}_{n}\v{S}_{n+1})
+J_{\tau}
(\v{\tau}_{n}\v{\tau}_{n+1}+\varepsilon \tau_{n}^{z}\tau_{n+1}^{z})
\big\} \,,
\end{eqnarray}
where $\v{S}_{n}$ and $\v{\tau}_{n}$ are respectively the
spin-$1\over2$ and pseudospin-$1\over2$ operators at the site $n$. The
Hamiltonian (\ref{ham}) can be viewed as two spin-$1\over2$ chains
coupled by a biquadratic interaction only (a generalized spin ladder).

Generally, the model has an $\rm SU(2)\times U(1)$ symmetry, so that
the total spin $(S,S^{z})$ of the $\v{S}$-chain, as well as the
$z$-projection $\tau^{z}$ of the total spin of the $\v\tau$-chain are
good quantum numbers.  Under certain simplifying assumptions
(neglecting Hund's rule coupling, nearest neighbor hopping between
the same type of orbitals only, and only one Coulomb on-site repulsion
constant), one obtains the above Hamiltonian with
$\varepsilon=\varepsilon'=0$, $J_{s}=J_{\tau}={1\over4}K$ which
possesses the hidden SU(4) symmetry. \cite{Li+98,Ueda+98} At this
special point, the model is Bethe ansatz solvable \cite{ULS} and
gapless. Depending on the microscopic details of the interaction, this
high symmetry can be broken in several ways. For example, finite Hund's rule
coupling or existence of more than one Coulomb repulsion constant
makes $J_{s}$ and $J_{\tau}$ different, and local crystal fields can
induce considerable anisotropy in the orbital sector.  

The phase diagram of the isotropic ($\varepsilon=\varepsilon'=0$) $\rm
SU(2)\times SU(2)$ version of the model (\ref{ham}) in the vicinity of
the $SU(4)$ point was analytically studied recently by several authors.
\cite{Azaria+99,Affleck+99,LeeLee00}
Extensive density matrix renormalization group (DMRG) studies of the
phase diagram \cite{Affleck+99,Pati+98,Ueda+99} established the form
of phase boundaries and the quantum numbers of the lowest-lying
excitations.\cite{Ueda+99}

Moving away off the $SU(4)$ point towards larger $J_s$, $J_\tau$, one
runs into the spontaneously dimerized phase  with a
finite gap and twofold degenerate ground state. The 
structure of the
elementary excitation spectrum in the dimerized phase is known rather
incompletely. The weak coupling
($J_{s,\tau}\gg K$) region
of the dimerized phase is known to be a realization of the
so-called non-Haldane spin liquid \cite{NT97} where magnons become
incoherent excitations since they are unstable against the decay into
soliton-antisoliton pairs. Variational study for a special ``dimer point''
$J_{s}=J_{\tau}={3\over4}K$, where the exact ground state in a form of
a simple valence bond solid state is known, \cite{KM98} confirms that
solitons are the lowest excited states also in case of strong
coupling. However, recent DMRG results \cite{Ueda+99} revealed a
complicated pattern of the change in quantum numbers of the
lowest-lying excited states in the vicinity of this special point;
the standard DMRG technique provides only the
information about the quantum numbers and energy of the lowest excited
state, but not the information on the entire spectrum.  Recent 
work \cite{YuHaas00} using the so-called dynamical DMRG method contains data
for the dynamical structure factor, but  
for only one point in the dimerized phase, namely for the dimer point. 

Further, realistic material-relevant models are generically anisotropic
($\varepsilon,\varepsilon'\not=0$) and asymmetric ($J_{s}\not=J_{\tau}$);
\cite{Kugel-Khomskii,Sa-Gros00} the anisotropic model of the same type
arises also as an effective Hamiltonian for a spin tube.\cite{Orignac+}
Thus we think it is of interest to get insight into the properties of
the elementary excitations in the dimerized phase of the more general
asymmetric/anisotropic model, which is the aim of the present paper.

In section \ref{sec:exact} we introduce briefly the ``dimer line'' in
the space of model parameters where the exact twofold degenerate
spontaneously dimerized ground state is known. Based on this picture
of the ground state, in section \ref{sec:ee} we present the
variational study of the interplay between different types of
elementary excitations, namely, between soliton-antisoliton pairs and
their bound states (magnons). We show that soliton pairs are the
lowest excitations only in a narrow region of the phase diagram, and
away from the dimer line the low-energy physics is determined by
magnon-type localized excitations which are essentially
soliton-antisoliton bound states.  
An additional phase boundary is shown to appear when the anisotropy
exceeds a certain critical value.
Variational dispersion curves are
found to agree qualitatively with the exact diagonalization (Lanczos)
results in the vicinity of the dimer line.

In section \ref{sec:holes} we describe some exact results for the
doped spin-orbital model on the ``dimer line''. We show that the
natural charge excitation in this case is a ``charged soliton'', i.e.,
the bound state of a hole and a soliton in the dimer order.  It is
found that under certain condition ``charged solitons'' themselves
exhibit a tendency to form bound states; exact expressions for the
energy of those bound states are presented.

\section{Exact ground states}
\label{sec:exact}

\begin{figure}
\begin{center}
\mbox{\psfig{figure=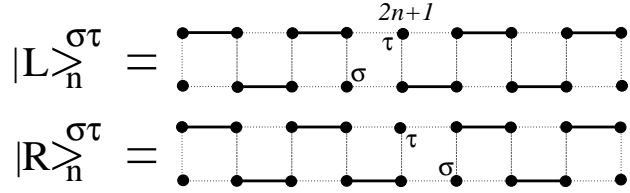,width=82mm,angle=0}}
\end{center}
\vspace{3mm}
\caption{\label{fig:ee} The single soliton states
$|L\rangle_{n}^{\sigma\tau}$, $|R\rangle_{n}^{\sigma\tau}$
used in Eq.\
(\protect\ref{sol}). The lower and the upper chains correspond to real
spins ($\v S$) and orbital pseudospins ($\v \tau$),
respectively. Thick solid lines denote singlet bonds. The soliton
connects two degenerate spontaneously dimerized states which are exact
ground states of the model (\protect\ref{ham1}) at $J_{\tau}=3/4$,
$J_{s}=(3+\varepsilon)/4$.}
\end{figure}

In addition to the $SU(4)$ model, there are several other particular
points in the phase diagram of (\ref{ham}), for which exact results
are available.\cite{KM98,Itoh99,Martins-Nienhuis00}  Particularly, at
the point $J_{s}=J_{\tau}=3K/4$ the exact ground state \cite{KM98} is
a product of checkerboard-ordered spin and orbital singlets (see Fig.\
\ref{fig:ee}), so that the spin and orbital degrees of freedom are
completely decoupled in the ground state. This ground state is
spontaneously dimerized and thus twofold degenerate in case of
periodic boundary conditions (for open boundaries the degeneracy would
be 8-fold).

One can show \cite{Itoh99} 
that the same ground state persists in a finite range
of anisotropy in the case of the special choice
$\varepsilon=\varepsilon'$. In this case, up to a constant term, the
model (\ref{ham}) can be recast in the form
\begin{equation} 
\label{ham1} 
\widehat{H}=\sum_{n}(\v{S}_{n}\v{S}_{n+1}+J_{\tau})
(\v{\tau}_{n}\v{\tau}_{n+1}+\varepsilon
\tau_{n}^{z}\tau_{n+1}^{z}+J_{s})\,, 
\end{equation} 
where we have set $K$ to be the unit energy scale. Then it is easy to
see that the choice $J_{\tau}=3/4$, $J_{s}=(3+\varepsilon)/4$ makes
the checkerboard-type singlet product wave function an eigenstate, and
one can prove that it stays the ground state as long as $-2
<\varepsilon<\infty$. Thus, in the 3D phase space
$(J_{s},J_{\tau},\varepsilon)$ one has a line with an exactly known
ground state; in what follows we will call it the ``dimer line'' for
the sake of brevity.  This line ends at the multicritical point
$\varepsilon=-2$, where the ground state becomes degenerate with
infinitely many other states: e.g., changing any of the orbital
($\v{\tau}$) singlets into a triplet with $\tau^{z}=\pm1$ does not
change the energy.  The point $\varepsilon=+\infty$ is another
multicritical point where the triplets with $\tau^{z}=0$ can be
created with no energy cost. For $\varepsilon<-2$ the ground state is
also exactly known \cite{Itoh99} and is a product of the ferromagnetic
state in $\v \tau$-sector and the  antiferromagnetic Bethe-ansatz
state in the $\v S$-sector. The energy of ferromagnons softens at
$k=\pi$ when $\varepsilon$ tends to $-2$.

We have also checked that (\ref{ham1}) at $J_{\tau}=3/4$,
$J_{s}=(3+\varepsilon)/4$ is the only nontrivial model with exactly
known ground state (within the physically relevant subspace of
Hamiltonians defined by (\ref{ham})) which can be found within the
matrix-product based optimum ground states approach (see e.g. Ref.\
\onlinecite{KM98rev} and references therein), at least if one uses a
rather general matrix-product ansatz
\[
|\Psi\rangle=\mbox{Tr}(g_{1}(C)g_{2}(\widetilde{C})
\cdots g_{N-1}(C)g_{N}(\widetilde{C}) )
\]
where $C\equiv\{x,y,x',y',p,q\}$, and 
the elementary matrix $g_{n}(C)$ has the following form:
\[
g_{n}(C)=\left(
\begin{array}{lr} 
x|\uparrow\downarrow\rangle_{n} +y|\downarrow\uparrow\rangle_{n} &
-p|\downarrow\downarrow\rangle_{n} \cr
q|\uparrow\uparrow\rangle_{n} &
x'|\uparrow\downarrow\rangle_{n} +y'|\downarrow\uparrow\rangle_{n}
\end{array}
\right)\,.
\]
Here $n$ denotes the rung of the ``ladder,'' and the first and second
arrows denote the states of $\v S$ and $\v \tau$ spins, respectively.

\section{Variational study of the elementary excitations}
\label{sec:ee}

In the vicinity of the dimer line, the
elementary excitations can be studied variationally.
There are essentially two
types of excitations: (i) solitons  connecting the
two degenerate spontaneously dimerized ground states (exactly
speaking, soliton pairs), and (ii) magnons which are localized excitations
and can be viewed as soliton-antisoliton bound states. 
The variational results give an upper bound for the dispersion at
the dimer line where the ground state is exact, and one may expect that they
provide a reasonably good approximation in the vicinity of this line.

\subsection{Solitons}

For a single soliton excitations one can use the variational ansatz of
the type introduced in Ref.\ \onlinecite{KM98}:
\begin{equation} 
\label{sol} 
|\alpha; p\rangle^{\sigma\tau}=\sum_{n}\left\{ 
e^{ipx_{n}^{L}} |L\rangle_{n}^{\sigma\tau} 
+\alpha\,e^{ipx_{n}^{R}} |R\rangle_{n}^{\sigma\tau}\right\}\,.
\end{equation}
Here the states $|L\rangle_{n}^{\sigma\tau}$, $|R\rangle_{n}^{\sigma\tau}$ are
shown schematically in Fig.\ \ref{fig:ee}, and
$x_{n}^{L}=2n+{1\over2}$, $x_{n}^{R}=2n+{3\over2}$ are the effective
``center of mass'' coordinates of $|L\rangle$ and $|R\rangle$ states,
respectively, and   $\alpha$ is a variational parameter.
This soliton state has the quantum numbers
$S^{z}=\sigma$, $\tau^{z}=\tau$ and may be combined in pair with the
corresponding antisoliton state to get states with
$(S,\tau^{z})=(1,\pm1)$, $(0,0)$, $(1,0)$ or $(0,\pm1)$. Since there are no
diagonal exchange interactions in our ``ladder'', the variational
energy of a single soliton does not depend on $(\sigma,\tau)$ so that
we omit the corresponding index in what follows.

It is easy to see that there exist a couple of useful symmetry properties. 
Upon the operation of interchanging upper and lower leg
$(S\leftrightarrow \tau)$ the $R$ and $L$ soliton states transform as
$|L\rangle_{n}\mapsto|R\rangle_{n-1/2}$,
$|R\rangle_{n}\mapsto|L\rangle_{n+1/2}$, so that 
\begin{equation} 
\label{symST} 
|\alpha; p\rangle \mapsto \alpha|(\alpha)^{-1}; p\rangle\,.
\end{equation}
That means that if the variational energy of the state (\ref{sol})
calculated with the Hamiltonian (\ref{ham1}) has a minimum at a
certain $\alpha=\alpha_{0}$, then the energy of the same state
calculated with the Hamiltonian with interchanged $S$ and $\tau$
operators should have a minimum at $\alpha'=1/\alpha$. An important
consequence is that in the symmetric case $\varepsilon=0$,
$J_{s}=J_{\tau}$ the minimum should always occur at $\alpha=\pm1$. (It
should be remarked that this property was missed in our earlier
treatment;\cite{KM98} the variational parameter $\zeta\equiv \alpha
e^{ip}$ used in Ref.\ \onlinecite{KM98} was erroneously assumed to be
real, which forced the minimum to be at $\zeta=\pm1$).

Upon a reflection $x\mapsto -x$ the soliton states transform as
$|L\rangle_{n}\mapsto |R\rangle_{N-n}$, $|R\rangle_{n}\mapsto
|L\rangle_{N-n}$, so that one has
\begin{eqnarray} 
\label{symP} 
|\alpha; p\rangle &\mapsto& \alpha e^{ip(2N+1)}
|(\alpha)^{-1}; -p\rangle\nonumber\\
&=& \alpha e^{ip(2N+1)}
|(\alpha^{*})^{-1}; p\rangle^{*}\,. 
\end{eqnarray}
where the asterisk denotes complex conjugation.
The energies of two states with mutually conjugate wave functions should be
equal, therefore replacing $\alpha$ by $1/\alpha^{*}$ should
not change the variational energy, which means that the minimum
is reached on a subspace with $|\alpha|^{2}=1$. Thus, we can set
$\alpha=e^{i\varphi}$ in our variational ansatz (\ref{sol}).

Essentially simple but tedious calculation of the variational dispersion
yields 
\begin{eqnarray} 
\label{esol} 
E_{\rm sol}(p)=
{F_{0}(p) + F_{c}(p)\cos \varphi +F_{s}(p)\sin\varphi
\over 
24(25-16\cos^{2}p)(5+4\cos\varphi\cos p) }\,,
\end{eqnarray}
where we have used the notation
\begin{eqnarray} 
\label{fun}
F_{0}(p)&=&[J_{\tau}(\varepsilon+3)+3J_{s}](383-392\cos(2p))\nonumber\\
        &+&(\varepsilon+3)(48\cos(4p)+180\cos(2p)-93),\nonumber\\
F_{c}(p)&=&4\cos p (33-32\cos2p)
\{ (2J_{\tau}-3)(\varepsilon+3)+6J_{s}\},\nonumber\\
F_{s}(p)&=&20\sin p (25-16\cos^{2}p)\{J_{\tau}(\varepsilon+3) -3J_{s}\}\,.
\end{eqnarray}
It is now easy to minimize the expression (\ref{esol}) in $\varphi$
and to obtain the final variational soliton energy. The equation
$dE_{s}/d\varphi=0$ always has two roots, $\varphi_{0}$ and
$\pi-\varphi_{0}$, and one of them yields the minimum for
$0<p<{\pi\over2}$, and the other does so for ${\pi\over2}<p<\pi$. In
this way, though one cannot see that from Eq. (\ref{esol}) directly, the
final minimized soliton energy $E_{\rm sol}^{\rm min}(p)$ is symmetric with
respect to the shift $p\mapsto p+\pi$, in full agreement with the fact
that the Brillouin zone in the dimerized phase is halved.

For certain special cases the  variational 
soliton dispersion can be sufficiently simplified. It is easy to see that
under the following condition of {\em ``generalized symmetry'':}
\begin{equation} 
\label{gensym} 
J_{s}=J(1+\varepsilon/3),\qquad J_{\tau}=J
\end{equation}
the function $F_{s}(p)$ identically vanishes, so that the minimum of
(\ref{esol}) is reached at $\varphi=0$ or $\pi$ (i.e., $\alpha=\pm1$),
and one obtains
\begin{eqnarray} 
\label{Esymgen} 
E_{\rm sol}^{\rm sym}&=&(3+\varepsilon)/24 \nonumber\\
&\times& {62J+12\cos 2p +3
-(64J+12)|\cos p| \over 5-4|\cos p|}\,.
\end{eqnarray}
At the dimer line, i.e., for $J={3\over4}$, the above expression
simplifies further to
\begin{equation} 
\label{Edimer} 
E_{\rm sol}^{\rm dimer}={3+\varepsilon\over16}(5-4|\cos p|)\,.
\end{equation}
The energy gap at the dimer line occurs always at $p=0$, $\pi$. One
can check that the gap which follows from (\ref{Edimer}) coincides
with the result given in Ref.\ \onlinecite{KM98}; however, for general
$p$ the formula (\ref{Edimer}) yields the lower energy and is in
addition much simpler than the corresponding expression in Ref.\
\onlinecite{KM98}.

The energy of the continuum of soliton-antisoliton scattering states
is given by
\[
E_{pair}(k,q)=E_{s}(k/2+q)+E_{a}(k/2-q)\,,
\]
where $k$ and $q$ are respectively the total and relative momentum.
All momenta here are defined for the original non-dimerized chain, and
the Brillouin zone is in fact halved, so that $k\in[-\pi/2,\pi/2]$.

\subsection{Magnons}

\begin{figure}
\begin{center}
\mbox{\psfig{figure=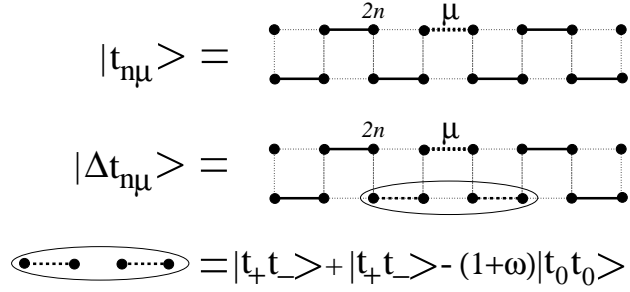,width=82mm,angle=0}}
\end{center}
\caption{\label{fig:magnon}
Magnon states used in Eqs.\ (\protect\ref{mag-mix}),
(\protect\ref{mag-ansatz}). Here $\omega=0$ for $\tau$-magnons (i.e.,
when the upper leg here is the $\tau$ chain) and $\omega=\varepsilon$ for
$S$-magnons (the upper leg is the $S$ chain).  
}
\end{figure}

We understand ``magnons'' as localized excitations which do not
break the dimer order (in contrast to solitons).  In a crude
approximation, one could imagine that a magnon state corresponds to exciting
one of the singlet bonds on the $S$ or $\tau$ chains to a triplet and
making it propagate along the chain. 
One may speak about $S$- or
$\tau$-magnons, respectively.
In turns out, however, that such a crude single-mode approximation
overestimates considerably the magnon energy. A simple reason is that
a state $|t_{n\mu}\rangle$ shown in Fig.\ \ref{fig:magnon} is not an
eigenstate of the Hamiltonian even at the dimer line, and under the
action of the Hamiltonian it gets mixed with other states. In the
vicinity of the dimer line the largest contribution to the first-order
matrix elements (i.e., of the
$\langle\psi|\widehat{H}|t_{n\mu}\rangle$ type) comes from the
action of the local Hamiltonian $\widehat{h}_{n}$ coupling the sites $(2n+1)$
and $(2n+2)$ (see Fig.\ \ref{fig:magnon}). 
The other first-order contributions  are proportional to small parameters
$\lambda_\tau=J_{\tau}-{3\over4}$,
$\lambda_s=J_{s}-{3+\varepsilon\over4}$,
measuring the deviations from the dimer line, and we neglect them for
the sake of consistency (see below). Thus, it
looks reasonable to take a linear combination of the two orthogonal
states
$|t_{n\mu}\rangle$ and  
\begin{equation} 
\label{mag-mix} 
|\Delta^{(1)} t_{n\mu}\rangle=\widehat{h}_{n}|t_{n\mu}\rangle-
\langle t_{n\mu}|\widehat{h}_{n}|t_{n\mu}\rangle
\langle t_{n\mu}|
\end{equation}
(see Fig.\ \ref{fig:magnon}) as a simplest variational ansatz. The
magnon state with momentum $k$ can be written as
\begin{equation} 
\label{mag-ansatz}
|m;k\rangle =\sum_{n}e^{ik(2n+1)}\left\{
\cos\chi |t_{n\mu}\rangle +\sin\chi |\Delta^{(1)} t_{n\mu}\rangle\right\}\,.
\end{equation}
where $|\Delta^{(1)} t_{n\mu}\rangle$ is assumed to be normalized, and $\chi$
is a variational parameter. After minimization in $\chi$, the magnon
dispersions take the following form:
\begin{eqnarray}  
\label{magnons}    
E_{S}(k)&=& e_{0,S}+(1/8)(3+\varepsilon-4J_{s})\cos(2k)\,, \nonumber\\
E_{\tau}^{0}(k)&=& e_{0,\tau}^{0}
+(1/8)(1+\varepsilon)(3-4J_{\tau})\cos(2k)\,,\\
E_{\tau}^{\pm1}(k)&=&
e_{0,\tau}^{\pm1}+(1/8)(3-4J_{\tau})\cos(2k)\,,
\nonumber
\end{eqnarray}
where the superscripts for $\tau$-magnons indicate the value of
$\tau^{z}$. One may observe that the inclusion of the above-mentioned
contributions to $|\Delta^{(1)} t_{n\mu}\rangle$ which are
proportional to $\lambda_{s}$, $\lambda_{\tau}$ would lead to
corrections of the order $O(\lambda_{s,\tau}^{2})$ in both the
bandwidth and the self-energy of magnons. To be consistent, we have to
omit those contributions, since we have already implicitly neglected
the corrections to the ground state energy which are also
$O(\lambda_{s,\tau}^{2})$ in the leading order.  (This does {\em
not\/} mean, of course, that our calculation will have the accuracy
$O(\lambda_{s,\tau}^{2})$, since the contributions from the
higher-order matrix elements of the type $\langle
\psi|(\widehat{H})^{m}|t_{n\mu}\rangle$ do not contain any small
parameter.  The ansatz (\ref{mag-ansatz}) should be considered as
purely variational, sort of ``one-step Lanczos'' trick).

The self-energies for $\tau$-magnons can be
compactly written as
\begin{eqnarray}
\label{taumag} 
e_{0,\tau}^{\mu}&=&3J_{s}/4+J_{\tau}\nonumber\\
&-&{(1+\eta \varepsilon) +2\big\{ (1+\eta\varepsilon)^{2}
+48J_{s}^{2} \big\}^{1/2} \over 16},
\end{eqnarray}
where $\eta=1$ for $\mu=\pm1$ and $\eta=-1$ for $\mu=0$.
The expression for the self-energy of an $S$-magnon is somewhat cumbersome
due to the fact that the state $|\Delta^{(1)} t_{n\mu}\rangle$ in case of
the $S$-magnons does not have a certain value of the total orbital
momentum $T$ and is a mixture of states with $T=0$
and $2$:
\begin{eqnarray} 
\label{smag}
e_{0,S}&=&J_{s}+J_{\tau}
-{4J_{\tau}(3-\varepsilon)+3(1+\varepsilon) +\sqrt{Z}\over
16(\varepsilon^{2}+2\varepsilon+3)} ,\nonumber\\
Z&=&
16J_{\tau}^{2}(\varepsilon^{2}+2\varepsilon+2)
[\varepsilon^{4}+4\varepsilon^{3}+27(\varepsilon^{2}+2\varepsilon+2)]
\nonumber\\ 
&+&
8J_{\tau}\varepsilon^{2}
(\varepsilon^{4}+6\varepsilon^{3}+21\varepsilon^{2}+32\varepsilon+24)
\nonumber\\  
&+&
\varepsilon^{6}+6\varepsilon^{5}+21\varepsilon^{4}+44\varepsilon^{3} 
+36(2\varepsilon^{2}+2\varepsilon+1)
\end{eqnarray}

One can observe that the ansatz for a magnon state which we
have used above is in fact equivalent to the simplest possible ansatz for
a bound state of soliton and antisoliton, where only the configurations
of the type
\[
|L\rangle_n|\bar L\rangle_{n+{1\over2}},\quad   
|R\rangle_n|\bar R\rangle_{n+{1\over2}},
\quad |L\rangle_n |\bar R\rangle_{n+{1\over2}},\quad 
|R\rangle_n|\bar L\rangle_{n+{1\over2}}
\]
with soliton and antisoliton being nearest neighbors are taken into
account. Here the bars denote antisoliton states; note that for fixed
boundary conditions, solitons and antisolitons live on different
sublattices, e.g. the free spin in the $\v\tau$ chain (see Fig.\
\ref{fig:ee}) is placed on the odd (even) rung for solitons
(antisolitons), respectively, which is reflected in the above
notation.

This indicates that magnons should be viewed as bound
soliton-antisoliton states, rather than as independent
excitations. Our simple ansatz can capture only the short-range
physics of the tightly bound states with quantum numbers
$(S,\tau^{z})=(1,0)$, $(0,0)$, and $(0,\pm1)$; it will miss, however,
possibly existing higher-lying loosely bound states as well as states
with $(S,\tau^{z})=(1,\pm1)$.

It is interesting to remark that solitons themselves can be
viewed as bound states of two spinons belonging to the $S$ and $\tau$
chains, respectively, which should unbind when one moves towards the
exactly solvable $SU(4)$ point. Therefore on the way from the $SU(4)$
point to the dimer line there should exist an
interesting hierarchy of bound states, which would be an interesting
subject of the future studies.

\subsection{Crossover between solitons and magnons}

\begin{figure}
\begin{center}
\mbox{\psfig{figure=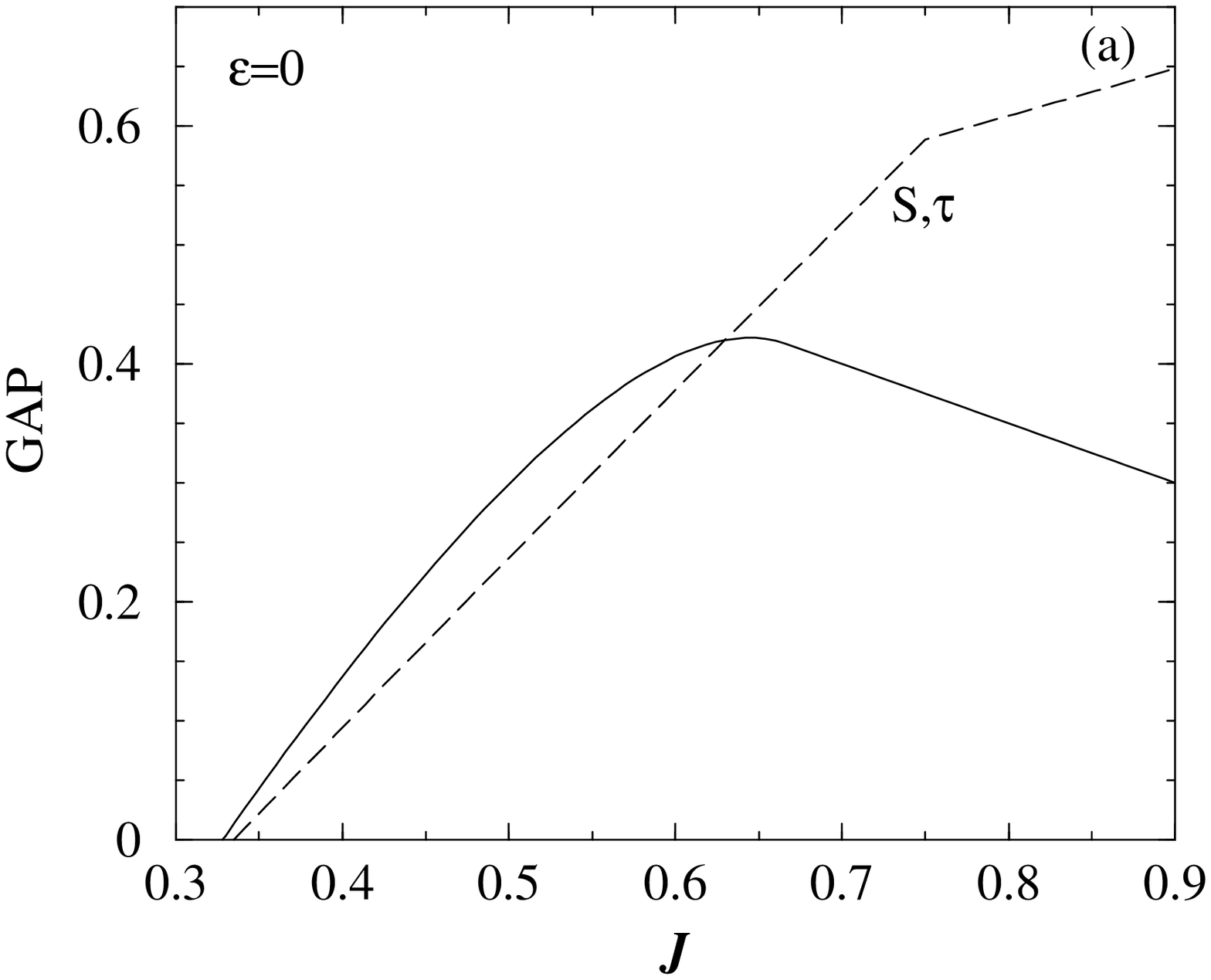,width=82mm,angle=0}}
\mbox{\psfig{figure=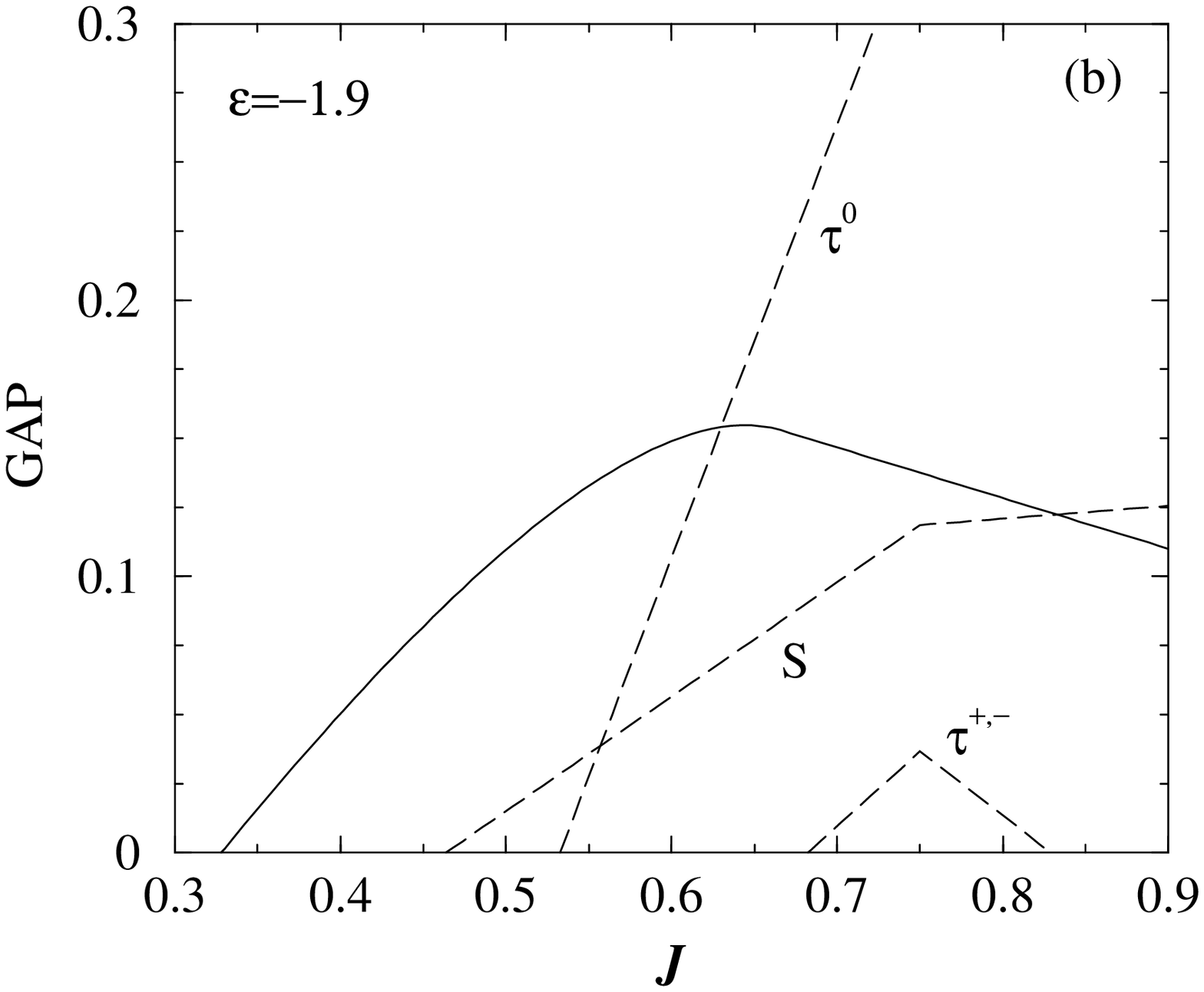,width=82mm,angle=0}} \\
\mbox{\psfig{figure=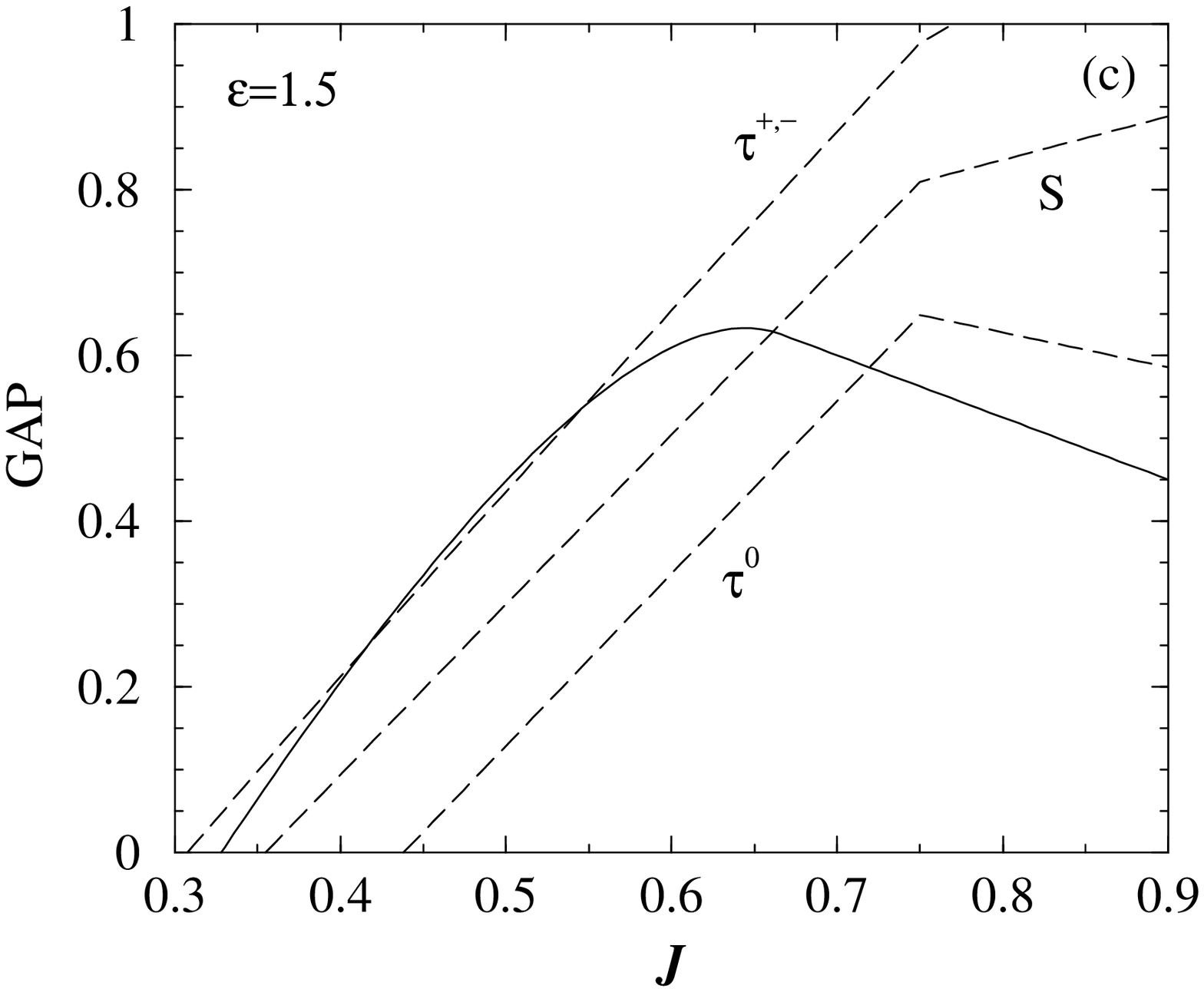,width=82mm,angle=0}}
\end{center}
\caption{\label{fig:gap} Variational results for the behavior of the
gaps for different types of excitations, along the ``generalized
symmetry line'' $J_{s}=J(1+\varepsilon/3)$, $J_{\tau}=J$, for a few
values of the anisotropy $\varepsilon$.  Magnon gaps (shown by dashed
lines) are straight lines. The gap of soliton-antisoliton pair
excitations (shown by a solid line) is a straight line while the
minimum of the soliton dispersion lies at $k=0$, and it starts to
deviate from a straight line for $J$ below certain threshold value,
when the minimum occurs at an incommensurate value of $k$.  }
\end{figure}

The minimum of the lower boundary of the soliton continuum is at $k=0$
in the immediate vicinity of the dimer line.  When $J_{s}$, $J_{\tau}$
decrease below certain values, the second equivalent minimum splits
off and starts to move towards $\pi/2$.  The magnon gap always lies
either at $k=\pi/2$ or at $k=0$.  Fig.\ \ref{fig:gap} illustrates the
behavior of the gap along the ``generalized symmetric line''
$J_{s}=J(1+\varepsilon/3)$, $J_{\tau}=J$ for a few choices of
$\varepsilon$.

One can observe the following important feature: the solitons are the
lowest excitations only in a limited region of the phase space. Even
at the dimer line line solitons cease to be the lowest excited states
if the anisotropy is sufficiently strong: the $\tau^{\pm}$-magnons
become the lowest excitations for $\varepsilon< -59/33$ and soften at
the phase transition point $\varepsilon=-2$ (respectively,
$\tau^{0}$-magnons become the lowest excitations for
$\varepsilon>59/7$ and the corresponding gap saturates at the finite
value of ${3\over4}$ at $\varepsilon\to+\infty$).  One can
see that whenever the variational gap closes while moving from the
dimer line, the magnon gap touches zero first, so that the phase
transitions are always governed by magnons.  

Three branches of $\tau$-magnons are degenerate at $\varepsilon=0$,
and for $\varepsilon>0$ ($\varepsilon<0$) the $\tau^{0}$-magnons
($\tau^{\pm}$-magnons) become lower in energy, respectively; the
$S$-branch lies always in between the $\tau^{\pm}$ and $\tau^{0}$
branches. In this connection, we would like to make a remark
concerning Fig.\ 4 of Ref.\ \onlinecite{PatiSingh00} which shows the
behavior of the $S$-magnon gap (the lowest excitation in the $(S^{z},
\tau^{z})=(1,0)$ sector) across one of the phase boundaries in the
model (\ref{ham1}) with $\varepsilon=-1$. We think the behavior of
$S$-gap is not relevant as a proof of the first order transition,
because another excitation, namely that of the $(0,1)$ sector, should
be responsible for the transition.

Away from the dimer line our description should become progressively worse
since the pure dimer state is too far from the correct ground
state. Thus the present approach cannot be used for estimation of the
shape of the phase boundaries, except when they become close to the
dimer line (see below).

Another observation is that though the gap of the soliton continuum
can be at incommensurate $k$, the lowest excited state almost always
(actually except for very narrow regions, whose existence may well be
an artifact of our approximation) occurs at the commensurate values of
$k=0$ or $\pi/2$ (see Fig.\ \ref{fig:gap}), which is consistent with
the results of Refs.\ \onlinecite{Azaria+99,Ueda+99}.  However, the
contribution from solitons can be observed in the {\em dynamic\/}
structure factor $S(k,\omega)$. Unfortunately, the only existing
dynamical DMRG study \cite{YuHaas00} provides data only for one point
in the dimerized phase (namely, for the point $J_{s}={3\over4}$,
$J_{\tau}={3\over4}$, $\varepsilon=0$) where no incommensurability is
expected.

One can see that if $\varepsilon$ tends towards the
multicritical points $\varepsilon=-2$ or $\varepsilon=+\infty$,
then there are {\em two\/} transitions along the generalized symmetry
line in the vicinity of the point
$J={3\over4}$ belonging to the dimer line: one at $J<{3\over4}$, and
another at $J>{3\over4}$. This indicates that if the anisotropy exceeds
a certain critical value, an 
additional transition boundary should appear in the
($J_{\tau}$, $J_{s}$) plane. Note that in this case the variational
phase boundaries lie close to the dimer line, so that this statement
can be considered reliable.

\subsection{Comparison with numerical results}

We have studied the spectrum of low-lying excitations of the model
(\ref{ham1}) by means of the exact diagonalization (Lanczos) method,
using a chain consisting of 12 sites (i.e. having 12 spin and 12
pseudospin degrees of freedom). 

\begin{figure}
\begin{center}
\mbox{\psfig{figure=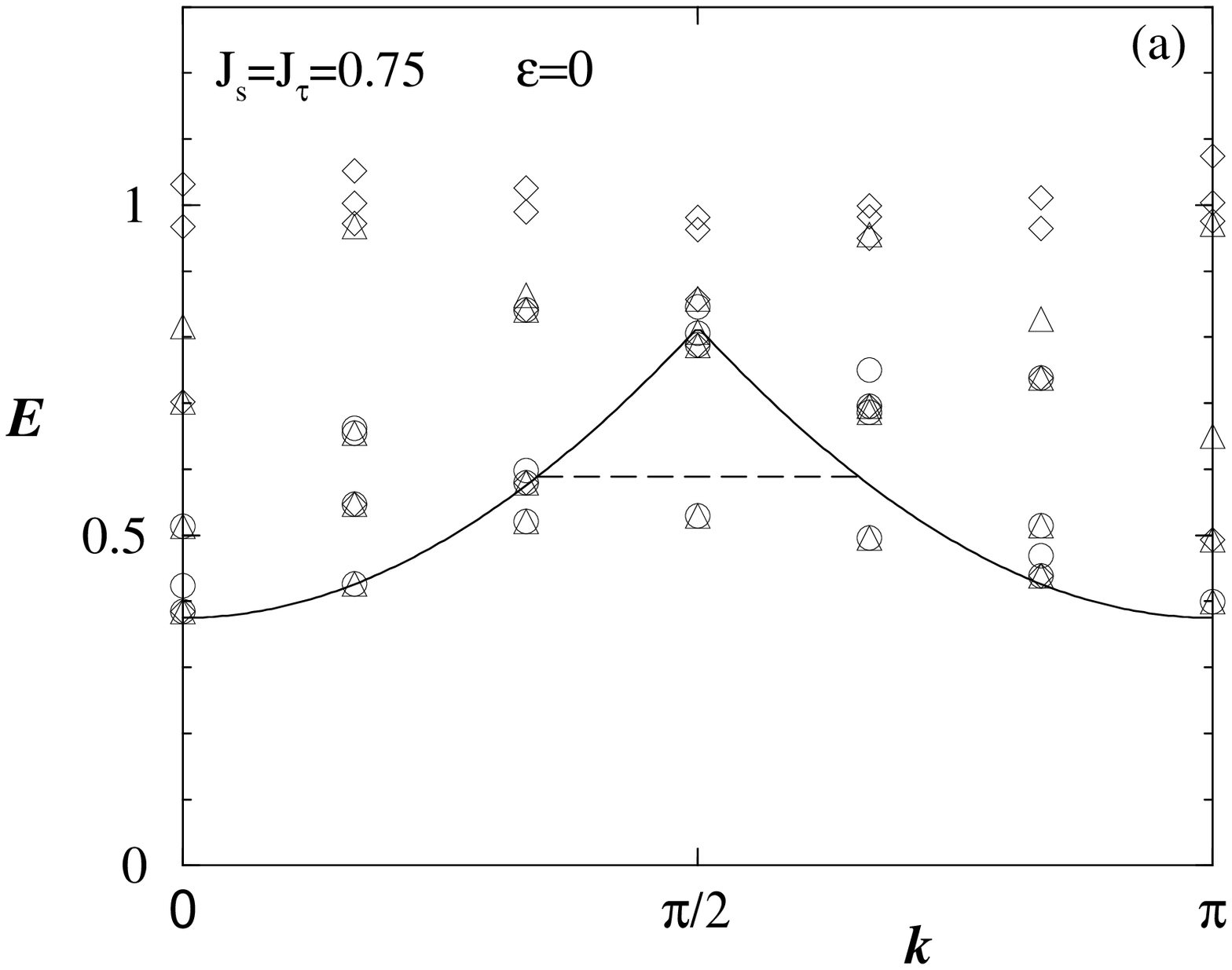,width=82mm,angle=0}}
\mbox{\psfig{figure=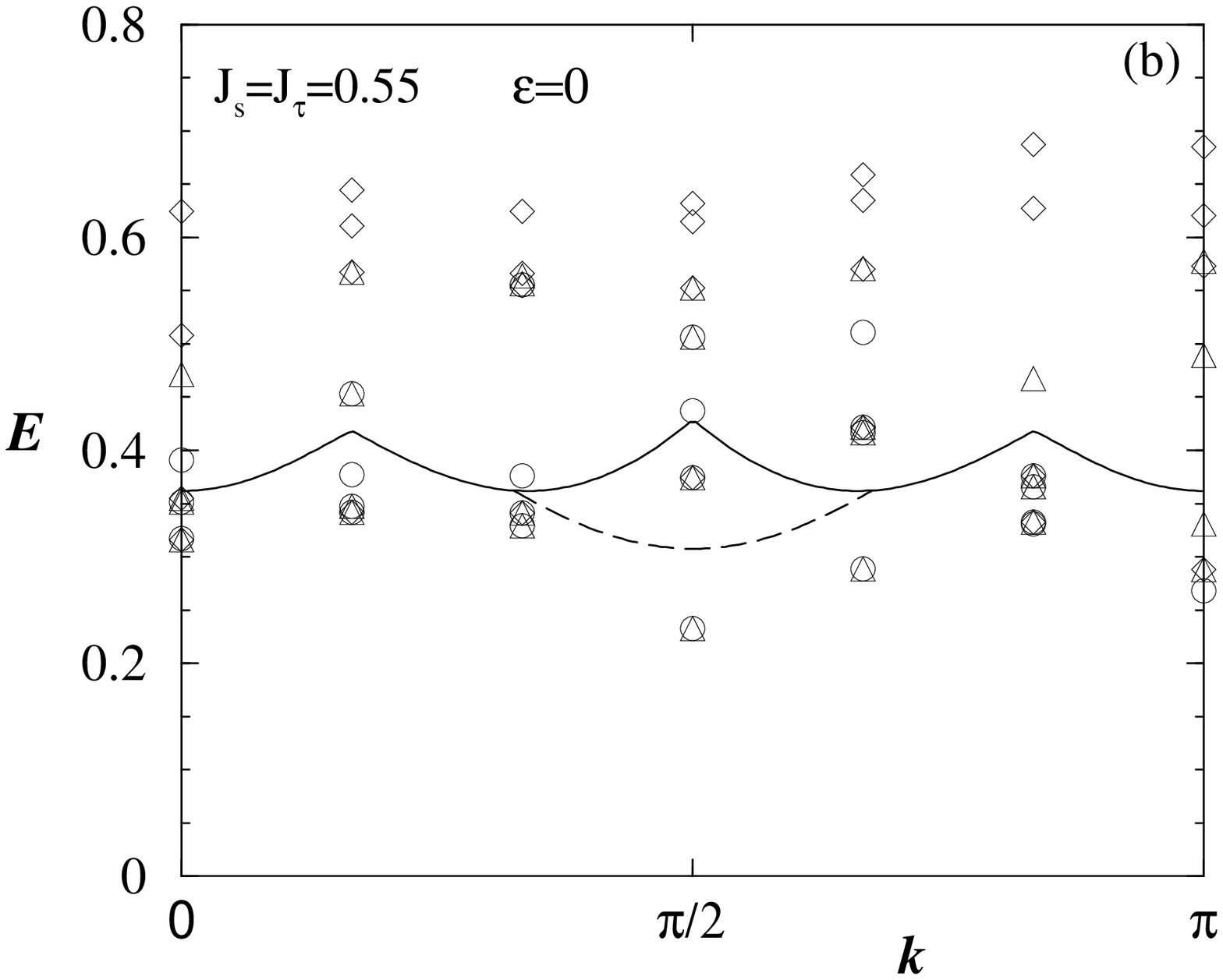,width=82mm,angle=0}} \\
\mbox{\psfig{figure=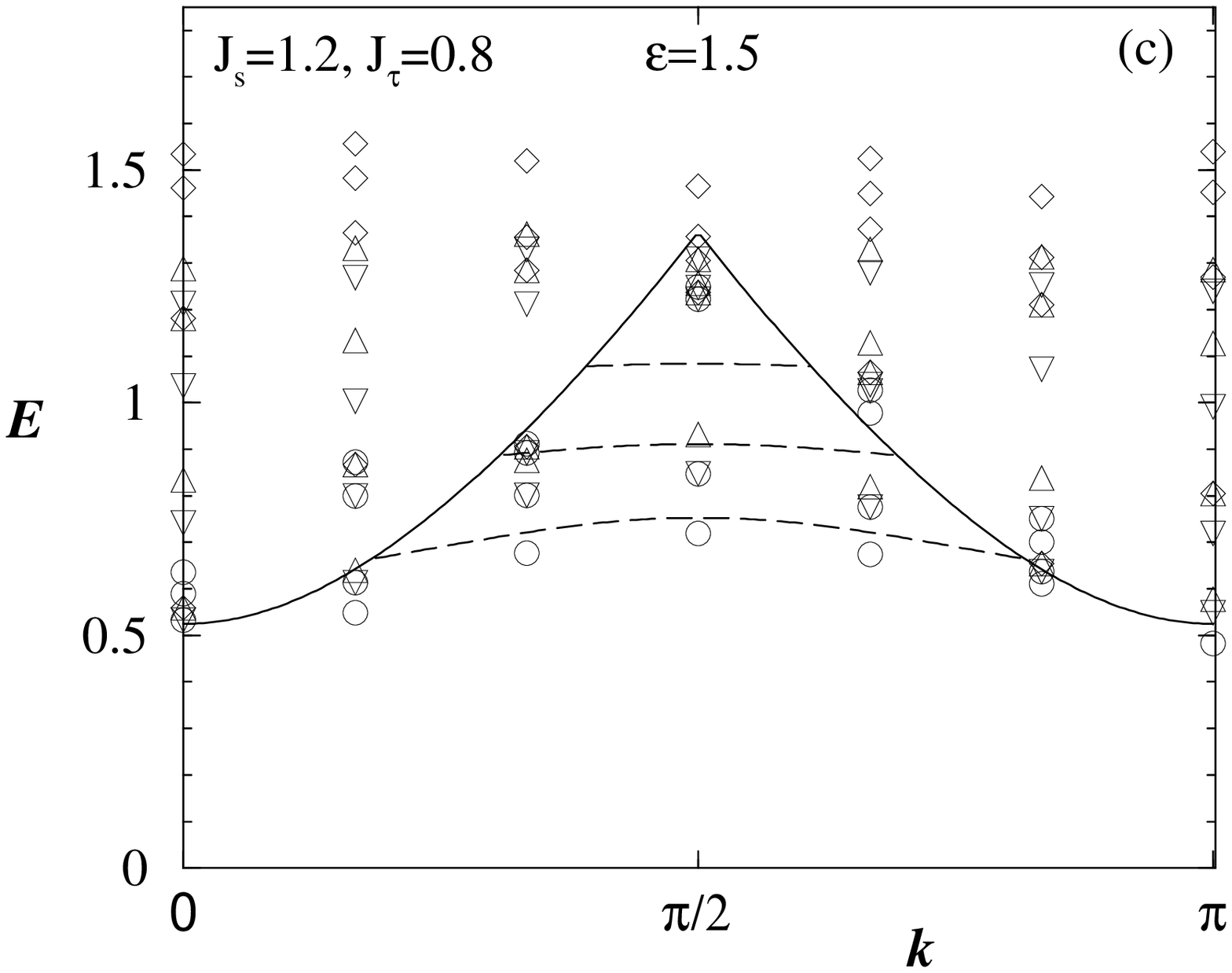,width=82mm,angle=0}}
\mbox{\psfig{figure=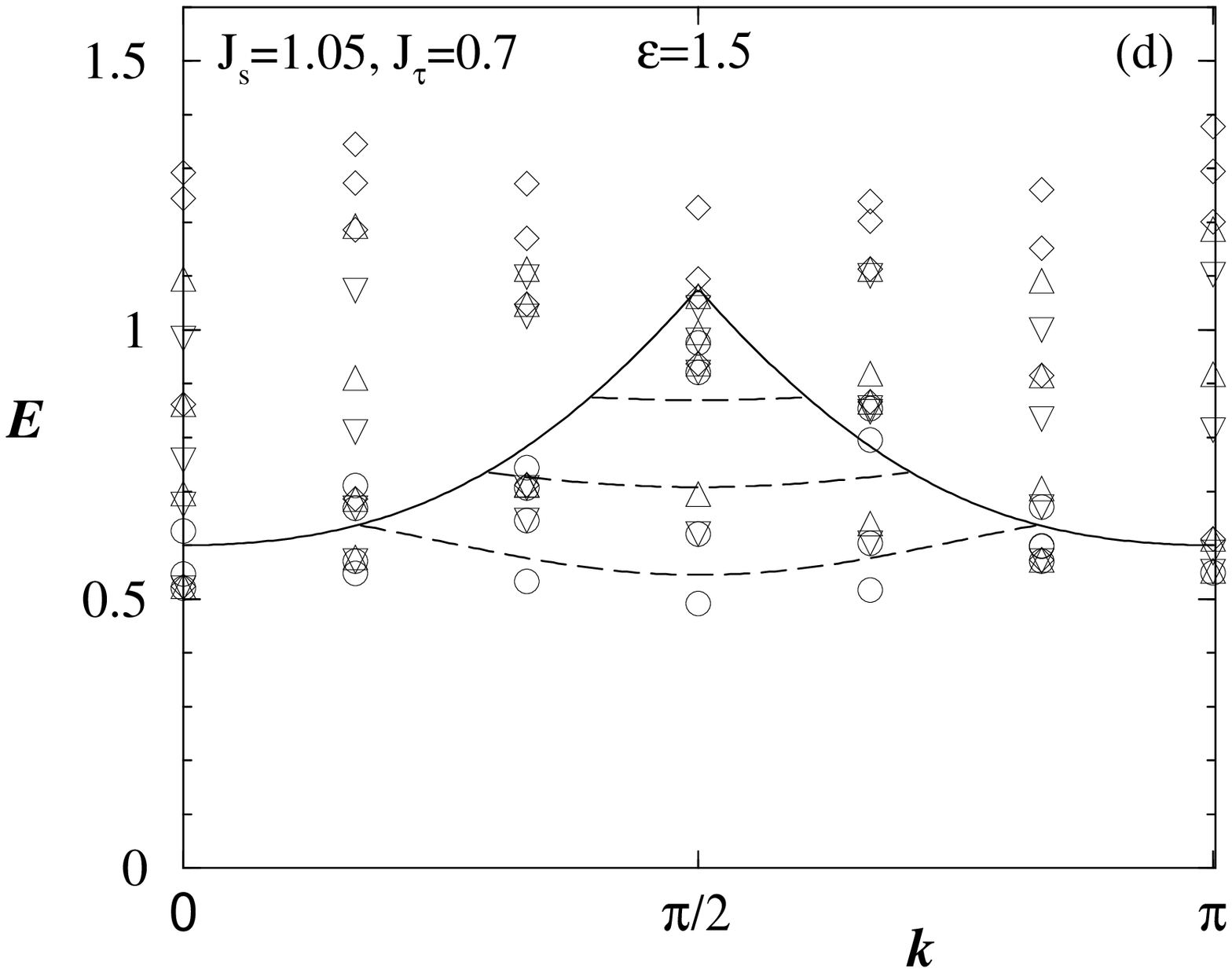,width=82mm,angle=0}} \\
\mbox{\psfig{figure=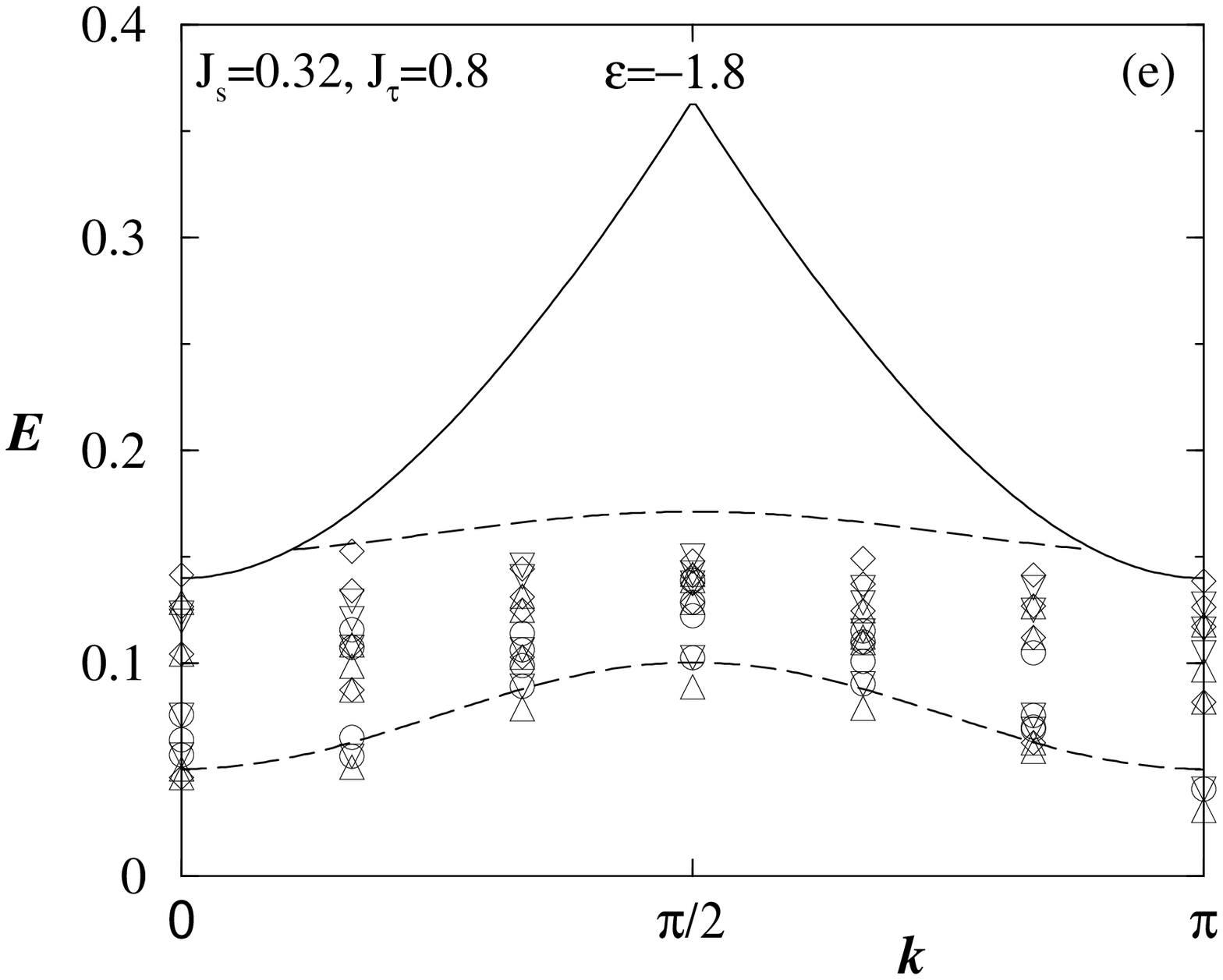,width=82mm,angle=0}}
\mbox{\psfig{figure=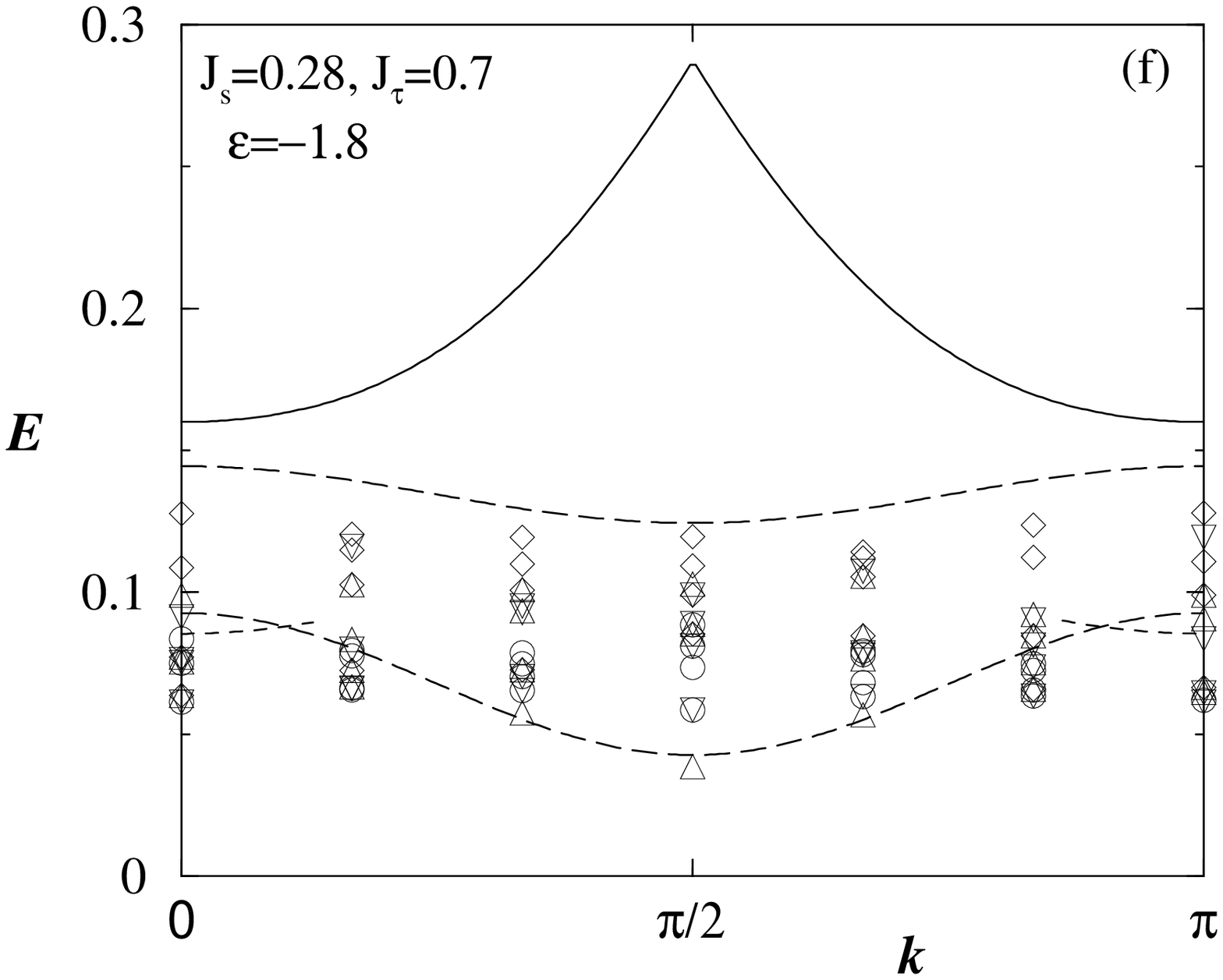,width=82mm,angle=0}}
\end{center}
\caption{\label{fig:disp} Behavior of the soliton and magnon
excitations along the line $J_{s}=J_{\tau}(1+\varepsilon/3)$ in the
$(J_{s},J_{\tau})$ plane for fixed $\varepsilon$. Solid lines indicate
the boundaries of the soliton continuum, and isolated long-dashed
lines correspond to the magnon-type excitations which can be viewed as
soliton-antisoliton bound states. In (f), additional dashed lines near
$k=0$ and $k=\pi$ show the lower boundary of the {\em magnon\/}
continuum formed by two $\tau^{+1}$ magnons.  Symbols denote the exact
diagonalization (Lanczos) data for a system of the length 12 (rungs);
circles, diamonds, up and down triangles correspond to the states with
the quantum numbers $(S^{z},\tau^{z})=(0,0)$, $(1,1)$, $(0,1)$, and
$(1,0)$, respectively. (For $\varepsilon=0$, up and down triangles are
degenerate and thus for the sake of simplicity, only up triangles are
displayed). }
\end{figure}

Fig.\ \ref{fig:disp} illustrates the behavior of the elementary
excitations spectrum for a few choices of $J_{s}$, $J_{\tau}$,
$\varepsilon$, in comparison with the variational results.  For each
value of the momentum $k$, the energies of four lowest-lying
eigenstates in each of the subspaces with $(S^{z},\tau^{z})=(0,0)$,
$(1,1)$, $(1,0)$, and $(0,1)$ were calculated. (The subspaces are
denoted in Fig.\ \ref{fig:disp} respectively with circles, diamonds,
down and up triangles).  One can see that our simple variational
ansatz is in a qualitative agreement with the Lanczos data for the
{\em lowest\/} excited states in the vicinity of the dimer line.  It
explains the main feature of the Lanczos data, namely, the rather flat
dispersions observed nearly everywhere in the dimerized phase.
One can also observe the tendency of solitons to have a spectrum with
the minimum at an incommensurate value of $k$ when $J_{s}$, $J_{\tau}$
decrease below certain value (see Fig.\ \ref{fig:disp}b).

Identification of different types of excitations can be done as follows:
Free soliton-antisoliton pairs can have any of the four
above-mentioned $(S^{z},\tau^{z})$ combinations, whose energy would be
exactly degenerate in the infinite system; that allows one to identify
the soliton states in Fig.\ \ref{fig:disp}a-d as highly degenerate (in
reality, nearly degenerate)
points where several different symbols are sitting on top of each
other. $S$-magnons appear as degenerate combination of circle and down
triangle, $\tau^{0}$-magnons should be circles, and $\tau^{+}$-magnons
must be up triangles.

The agreement concerning the higher-lying excitations is
considerably worse. Only for $\varepsilon=0$ (see
Fig. \ref{fig:disp}a,b) one can see the isolated magnon branch; for high
anisotropies in the vicinity of the multicritical point (Fig.\
\ref{fig:disp}e,f) a large number of states appears immediately above
the lowest-lying states appears, which may mean that actually {\em
many\/} bound states split from the soliton continuum, and we are able
to capture only the lowest-lying, most tightly bound states due to the
short-range nature of the magnon ansatz (\ref{mag-ansatz}).

\section{Doped spin-orbital model: charged solitons and their bound states}
\label{sec:holes}

It is interesting to study the behavior of the spin-orbital model
doped with holes (i.e., away from the quarter filling). In this case
one has to consider the corresponding $t$-$J$ model, adding the
appropriate hopping term to the Hamiltonian
$H(J_{s},J_{\tau},\varepsilon)$ determined by Eq.\ (\ref{ham1}):
\begin{equation} 
\label{tJ} 
\widehat{H}=\widehat{H}(J_{s},J_{\tau},\varepsilon) +
t\sum_{l\sigma\tau}(c^{\dag}_{l,\sigma\tau}c^{\phantom{\dag}}_{l+1,\sigma\tau}
+\mbox{h.c.})\,.
\end{equation}
Here $c_{l,\sigma\tau}$ are the electron Fermi-operators at the site
$l$, $\sigma$ and $\tau$ are the corresponding spin and orbital
quantum numbers, and the condition of no double occupancy is
implicitly assumed. We also assume, in a similar fashion as in the
course of derivation of (\ref{ham}), that hopping exists only between
the nearest neighbors with the same type of orbitals; the hopping
amplitude $t$ is chosen to be negative for the sake of
definiteness. One may also wish to add the nearest-neighbor Coulomb
repulsion term
\[
\widehat{H}\mapsto \widehat{H} 
+V\sum_{l}
\widehat{n}_{l}\widehat{n}_{l+1} \,,
\]
where
$\widehat{n}_{l}=\sum_{\sigma\tau}
c^{\dag}_{l,\sigma\tau}c^{\phantom{\dag}}_{l,\sigma\tau}$. 

\begin{figure}
\begin{center}
\mbox{\psfig{figure=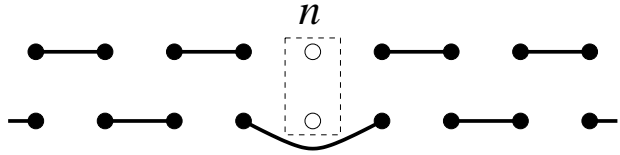,width=82mm,angle=0}}
\end{center}
\caption{\label{fig:hole} 
A ``charged soliton'' state $|h;n\rangle$ corresponding to a hole
localized on a soliton in the dimer order. }
\end{figure}

Consider a single hole in this system.
If one is on the dimer line $J_{\tau}=3/4$, $J_{s}=(3+\varepsilon)/4$,
where the checkerboard-type singlet
state is exact, it is easy to show that the hole state $|h;n\rangle$ shown
schematically in Fig.\ \ref{fig:hole} gets hopped to the neighboring
site without disturbing the structure of the ground state, so that the
corresponding translational invariant state with momentum $k$ is an
exact eigenstate with the energy 
\begin{equation} 
\label{1hole} 
E_{1h}(k)=2e_{0}+2t\cos(k),\qquad e_{0}\equiv {3(3+\varepsilon)\over16}\,.
\end{equation}
Such a state connects two regions with different dimerization pattern,
so that it can be viewed as a ``charged soliton'', or a hole bound on
a soliton. One can straightforwardly see that a hole state which does
not disturb the dimer order is not able to move and has to decay into
a ``charged soliton'' and an antisoliton.

\begin{figure}
\begin{center}
\mbox{\psfig{figure=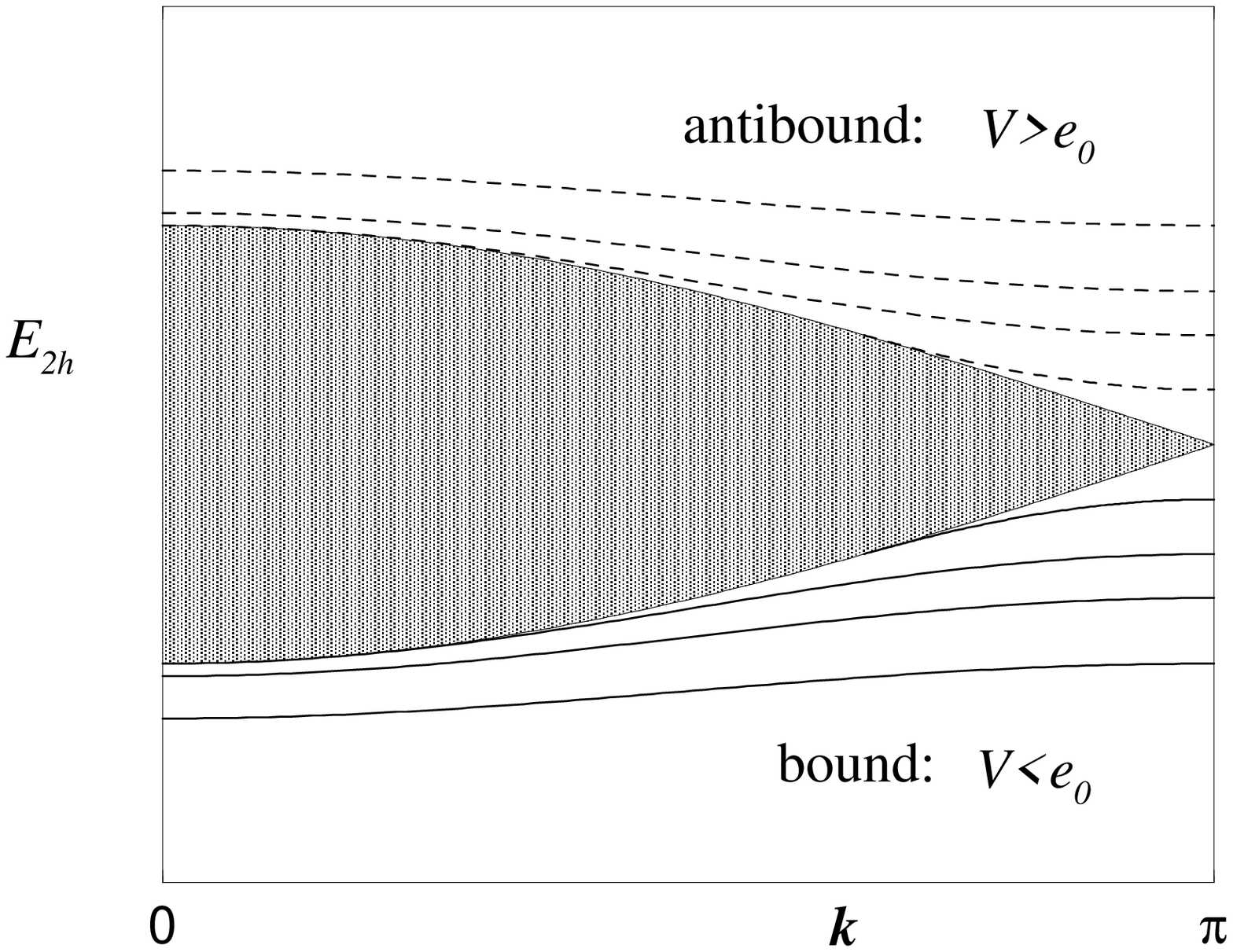,width=82mm,angle=0}}
\end{center}
\caption{\label{fig:2h} The spectrum of the two-hole problem, for
fixed $e_{0}$ and $t$, at several different values of $V$, as
discussed in the main text. Shaded
region denotes the continuum of scattering states formed by a
``charged soliton'' and ``charged antisoliton''; isolated solid and dashed
lines correspond to the bound and antibound states, respectively.}
\end{figure}

The two-hole problem can be also easily studied.
The general two-hole state with a total momentum $k$ can be written as
\begin{equation} 
\label{2hole} 
\Psi(k)=\sum_{n_{1}<n_{2}} e^{ik(n_{1}+n_{2})/2} f(n_{2}-n_{1})
|n_{1},n_{2}\rangle\,,
\end{equation}
and from the Schr\"odinger equation one obtains the following system
of equations for the function $f(r)$:
\begin{eqnarray} 
\label{f} 
&& 2t\cos{k\over2}\,[f(r+1)+f(r-1)]=(E_{2h}-4e_{0})f(r),\;\; r\geq 2\,,
\nonumber\\
&& 2t\cos{k\over2} \,f(2)=(E_{2h}-3e_{0}-V)\,f(1)\,.
\end{eqnarray}
Setting $f(r)=Ae^{iqr}+Be^{-iqr}$, with the parameter $q$ (having the
meaning of the relative momentum) being real, one obtains the
continuum of scattering states with the energy
\begin{equation} 
\label{2hc} 
E_{2h}(k,q)=4e_{0}+4t\cos(k/2)\cos(q/2)\,.
\end{equation}
There is also a bound state solution with $f(r)=f_{0}e^{-\kappa r}$;
the parameter $\kappa>0$ is determined by the equation
\begin{equation} 
\label{kappa}
e^{\kappa}={V-e_{0}\over 2t\cos(k/2) }\,.
\end{equation}
This solution describes a  localized state with the energy
\begin{equation} 
\label{2hb} 
E_{2hb}^{b}(k)=3e_{0}+V-{2t^{2}\over e_{0}-V}(1+\cos k)\,.
\end{equation}
The spectrum of the 2-hole problem is schematically shown in Fig.\
\ref{fig:2h}. 
One can see that for $V<e_{0}-2|t|$ the energy of this state lies
below the lower boundary of the two-hole continuum in the whole range
of $k$. When $e_{0}< V< e_{0}-2|t|$, the bound state exists only in a
certain interval $[k_{0},\pi]$ where $k_{0}$ is determined from
(\ref{kappa}) by demanding that $\kappa$ is positive; $k_{0}$ tends to
$\pi$ as $V$ tends to $e_{0}$. For $V>e_{0}$ the energy of the
localized state (\ref{2hb}) lies above the upper boundary of the
continuum, so that the states are ``antibound.''
Thus, the ``charged solitons,'' which are themselves
bound states of a hole and a soliton, tend, in their turn, to form
bound states, if the condition $e_{0}>V$ is satisfied.

One can observe that the general $N$-hole problem is equivalent to
that of the $XXZ$ ferromagnet with $J_{z}=e_{0}-V$ and $J_{xy}=2t$, 
in external magnetic field $h=e_{0}+V$ along the $z$
axis, and thus can
be treated by means of the Bethe ansatz. In fact, one can use the
known result for the energy of $N$-spin bound complex
\cite{Ovch67} 
to translate it into the corresponding formula for our problem:
\begin{eqnarray} 
\label{Nhole} 
E_{N}(k)&=&(e_{0}+V)N +\big\{ (e_{0}-V)^{2}-4t^{2}\big\}^{1/2}
\\
&\times& 
\Bigg\{ \tanh(N/N_{0})+ {2\over \sinh(2N/N_{0})}\sin^{2}(k/2) \Bigg\},
\nonumber
\end{eqnarray}
where $N_{0}$ is defined by the relation 
\[
\cosh(2/N_{0})=(V-e_{0})/2t\,.
\]
One can straightforwardly check that (\ref{2hb}) is the particular case
of (\ref{Nhole}) for $N=2$.  For large $N$ such a $N$-hole complex
corresponds to a classical soliton state in the $XXZ$ ferromagnetic
chain. \cite{KIK90} The expression (\ref{Nhole}) is valid for
$|V-e_{0}|>2|t|$. It is known \cite{Ovch67} that for  $|V-e_{0}|<2|t|$
only bound (respectively antibound) states
with $N=2$ exist, and the energies of states with larger $N$ fall into
the continuum. For $V-e_{0}<-2|t|$ the $N$-hole bound complex is
stable against the decay into smaller composite particles,
\cite{Ovch67,Gochev71} i.e.,
\[
E_{N}(k) < \min\{ E_{N-M}(k-k')+E_{M}(k') \} \,,
\]
and the opposite inequality (meaning respectively instability) is
fulfilled if $V-e_{0}>2|t|$.  Thus, for $V-e_{0}<-2|t|$ the ground
state of the model (\ref{tJ}) at finite doping represents a condensate
in which all holes are bound in a single ``particle'' with zero total
momentum;\cite{YangYang66} in the intermediate regime 
$-2|t|<V-e_{0}<2|t|$ only the
two-hole bound states are stable, and for $V-e_{0}>2|t|$ all bound
states become unstable.

\section{Summary}

We have studied, by means of the combination of variational and exact
results, as well as numerically, the interplay between different types
of elementary excitations in the model of an anisotropic spin-orbital
chain.  We analyze the vicinity of the special ``dimer line'' in the
phase space of the model where the spontaneously dimerized ground
state is known exactly.  Solitons in the dimer order are shown to be
the lowest excitations only in a narrow region of the phase diagram,
and the phase transitions are always governed by magnon-type
excitations which can be viewed as soliton-antisoliton bound states.
It is suggested that an additional phase boundary appears when the
anisotropy exceeds certain critical value.

We have considered also a $t$-$J$ version of the doped spin-orbital
model.  The elementary charge excitation is shown to be a ``charged
soliton'', or a soliton bound on a hole. We have studied the
conditions under which the ``charged solitons'' tend to form bound
states;    the exact expression for the energy of the $N$-hole bound
complex,  based on the known results from the Bethe Ansatz, is presented.

\acknowledgements

This work was initiated during the ``Quantum Magnetism'' conference at
the Institute for Theoretical Physics, Santa Barbara, and HJM wishes
to acknowledge the hospitality and support of the ITP during this
conference.  AK acknowledges the hospitality of Hannover Institute for
Theoretical Physics.  This work was supported by the German Federal
Ministry for Research and Technology (BMBFT) under the contract
03MI5HAN5.

\end{multicols}
\newpage

\newpage

\newpage

\newpage

\begin{references}

\bibitem[\ast]{leave} On leave of absence from the Institute of
Magnetism, 36(b) Vernadskii avenue, 03142 Kiev, Ukraine. 

\bibitem{NaTiSbO} E. Axtell, T. Ozawa, S. Kauzlarich, and
R.R.P. Singh, J. Solid State Chem. {\bf 134}, 423 (1997).

\bibitem{NaVO}M. Isobe and Y. Ueda, J. Phys. Soc. Jpn. {\bf 65}, 1178
(1996); Y. Fujii, H. Nakao, T.  Yosihama, M.Nishi, K. Nakajima,
K. Kakurai, M. Isobe, Y. Ueda, and H. Sawa, ibid. {\bf 66}, 326 (1997).

\bibitem{Kugel-Khomskii} I. Kugel and D. I. Khomskii, Sov. Phys. JETP {\bf
37}, 725 (1973); Sov. Phys. Usp. {\bf 25}, 231 (1982).

\bibitem{Li+98} Y. Q. Li, M. Ma, D. N. Shi, and F. C. Zhang,
Phys. Rev. Lett. {\bf 81}, 3527 (1998).

\bibitem{Ueda+98} Y. Yamashita, N. Shibata, and K. Ueda, Phys. Rev. B
{\bf 58}, 9114 (1998).


\bibitem{ULS} G. V. Uimin, JETP Lett. {\bf 12}, 225 (1970); C. K. Lai,
J. Math. Phys. {\bf 15}, 1675 (1974); B. Sutherland, Phys. Rev. B {\bf
12}, 3795 (1975).

\bibitem{Azaria+99} P. Azaria, A. O. Gogolin, P. Lecheminant, and
A. A. Nersesyan, Phys. Rev. Lett. {\bf 83}, 624 (1999);  P. Azaria, E.
Boulat, and P. Lecheminant, Phys. Rev. B {\bf 61}, 12112 (2000).

\bibitem{Affleck+99} C. Itoi, S. Qin, and I. Affleck, Phys. Rev. B
{\bf 61}, 6747 (2000).

\bibitem{LeeLee00} Yu-Li Lee and Yu-Wen Lee, Phys. Rev. B {\bf
61}, 6765 (2000).

\bibitem{Pati+98} S. K. Pati, R. R. P. Singh, and D. I. Khomskii,
Phys. Rev. Lett. {\bf 81}, 5406 (1998).

\bibitem{Ueda+99} Y. Yamashita, N. Shibata, and K. Ueda,
J. Phys. Soc. Jpn. {\bf 69}, 242 (2000).

\bibitem{NT97} A. A. Nersesyan and A. M. Tsvelik,
Phys. Rev. Lett. {\bf 78}, 3939 (1997).

\bibitem{KM98} A. K. Kolezhuk and H.-J. Mikeska, Phys. Rev. Lett. {\bf
80}, 2709 (1998).

\bibitem{YuHaas00} W. Yu and S. Haas, cond-mat/0005526.

\bibitem{Sa-Gros00} D. Sa and C. Gros, cond-mat/0004025.

\bibitem{Orignac+} E. Orignac, R. Citro, and N. Andrei,
Phys. Rev. B {\bf
61}, 11533 (2000).

\bibitem{Itoh99} K. Itoh, J. Phys. Soc. Jpn. {\bf 68}, 322 (1999).

\bibitem{Martins-Nienhuis00} M. J. Martins and B. Nienhuis,
cond-mat/0004238 (2000).

\bibitem{KM98rev} A. K. Kolezhuk and H.-J. Mikeska,
Int. J. Mod. Phys. B, {\bf 12}, 2325 (1998).

\bibitem{PatiSingh00} S. K. Pati and R. R. P. Singh, Phys. Rev. B {\bf
61}, 5868 (2000).


\bibitem{Ovch67} A. A. Ovchinnikov, Sov. Phys. JETP Lett. {\bf 5}, 38 (1967)

\bibitem{KIK90} A. M. Kosevich, B. A. Ivanov, and A. S. Kovalev,
Phys. Rep. {\bf 194}, 117 (1990).

\bibitem{Gochev71} I. G. Gochev, Sov. Phys. JETP {\bf 34}, 892 (1972).

\bibitem{YangYang66} C.N. Yang and C. P. Yang, Phys. Rev {\bf 151}, 258 (1966).





\end{references}
\end{document}